\documentclass[prl,twocolumn,tightenlines,showpacs,preprintnumbers,amsmath,amssymb]{revtex4}

\usepackage{graphicx}
\usepackage{dcolumn}
\usepackage{bm}
\usepackage{color}

\newcommand{\fref}[1]{Fig.~\ref{#1}}

\newcommand{\frefp}[2]{Fig.~\ref{#1}~(#2)}

\newcommand{\cref}[1]{chapter~\ref{#1}}

\newcommand{\Cref}[1]{Chapter~\ref{#1}}

\newcommand{\bref}[1]{(\ref{#1})}

\newcommand{\sub}[2]{{#1}_{\mbox{\!\! \scriptsize #2}}}

\newcommand{\bv}[1]{\mathbf{ #1 }}

\newcommand{\bra}[1]{\langle\,{#1}\, |}
\newcommand{\ket}[1]{|\,{#1}\,\rangle}

\usepackage{ulem}
\begin{document}

\preprint{}
\input{epsf.tex}

\epsfverbosetrue

\title{Van-der-Waals stabilized Rydberg aggregates}

\author{H.~Zoubi}
\email{zoubi@pks.mpg.de}
\affiliation{Max Planck Institute for the Physics of Complex Systems, N\"othnitzer Strasse 38, 01187 Dresden, Germany}
\author{A.~Eisfeld}
\affiliation{Max Planck Institute for the Physics of Complex Systems, N\"othnitzer Strasse 38, 01187 Dresden, Germany}
\author{S.~W\"uster}
\affiliation{Max Planck Institute for the Physics of Complex Systems,
  N\"othnitzer Strasse 38, 01187 Dresden, Germany}

\date{20 December, 2013}

\begin{abstract}
Assemblies of Rydberg atoms subject to resonant dipole-dipole interactions form Frenkel excitons.
We show that van-der-Waals shifts can significantly modify the exciton wave function, whenever atoms approach each other closely. As a result, attractive excitons and repulsive van-der-Waals interactions can be combined
to form stable one-dimensional atom chains, akin to bound aggregates. Here the van-der-Waals shifts ensure a stronger homogeneous delocalisation
of a single excitation over the whole chain, enabling it to bind up to six
atoms. When brought into unstable configurations, such Rydberg aggregates allow the direct monitoring of their dissociation dynamics.
\end{abstract}

\pacs{32.80.Ee, 37.10.Jk, 71.35.-y}

\maketitle

Van der Waals (vdW) forces between ground state molecules or atoms can lead to
the formation of molecular crystals and noble atom solids, without the need
for electron sharing between the individual constituents. Optical and
electrical properties in these aggregates are dominated by resonant
interactions of transition dipoles which lead to the appearance of Frenkel
excitons \cite{Fr30_198,frenkel_exciton}, in which an electronic excitation
can be delocalized in the lattice. Non-resonant interactions of vdW type lead to a change of the
transition energies, since they affect the excited state differently than the
ground state. This energy shift is homogeneous (i.e. the same for all monomers) for bulk crystals
\cite{Davydov,Agranovich,Zoubi2005} and in this context is termed gas to crystal
shift. For small structures of molecules, e.g~oligomers \cite{RoEiDv11_054907} or finite size domains on surfaces \cite{MPaMa13_064703,MPaMa13_044302}, the shift is not homogeneous any more.

Here we show that this inhomogeneity of the vdW shifts can strongly influence the entire exciton wave function.
To this end we consider assemblies of Rydberg atoms, which have huge transition dipoles (connecting two highly excited states) and hence can support Frenkel excitons over large distances \cite{RoHeTo04_042703,cenap:motion}.
Under conditions of one-dimensional confinement we further demonstrate the
possibility of self-assembled stable chains of Rydberg atoms, which form
Rydberg-``aggregates'' similar to the molecular situation \cite{Agranovich}. The
stable chain is formed by a competition between attractive forces generated by resonant dipole-dipole interactions and repulsive vdW interactions. This
bears some similarities to eximers. For short chains (e.g.$~$trimers) prepared in unstable configurations, we find an interesting break-up dynamics, reminiscent of molecular dissociation.
Due to the exaggerated properties of Rydberg atoms, the dissociation can be directly monitored in real space.

The basic effect of the inhomogeneous vdW shifts can be understood by the
following simple consideration. For comparatively large distances between
Rydberg atoms in a chain, resonant dipole-dipole processes of the type: $ns\ +\ np\ \leftrightarrow\ np\ +\ ns$ are
dominant \cite{Saffman,GallagherA} and support collective exciton states
\cite{RoHeTo04_042703,cenap:motion,cenap:emergingaggregate} which can have repulsive or attractive character \cite{cenap:motion}. If the distances between Rydberg
atoms in a chain become shorter, off-resonant contributions to the system's
electronic energies increase and can be modeled by the addition of vdW
potentials. This can give rise to on-site excitation energy shifts that
depend on the geometry of the atomic assembly. For a trimer, for example, this
effect can shift the central site out of resonance, since it has a
different local environment than the outer sites. In attractive exciton
states with repulsive vdW potentials, the site energy shifts near the chain center
cause a stronger delocalisation of the excitation towards the chain edges.
This results in much more homogeneous attraction throughout the chain, allowing the
stabilization of attractive exciton states
\cite{cenap:motion} by vdW repulsion. One-dimensional Rydberg chains can then essentially form bound
states with many atoms, resembling the vdW bound self assembled molecular
aggregates or molecular crystals.  Previous work focused on just two Rydberg atoms, that can form bound molecular states in three dimensions \cite{kiffner:dipdipdimer,kiffner:spinorbitcoupl,kiffner:nonabelian}.

For light Alkali atoms, atomic motion in the potentials discussed here can become relevant on the time-scale of Rydberg state lifetimes \cite{beterov:BBR}, resulting in flexible Rydberg aggregates \cite{cenap:motion,wuester:cradle,moebius:cradle,leonhardt:switching}. We investigate the behavior of such a flexible Rydberg trimer, and find that it may exhibit interesting dissociation dynamics that can be monitored in time and space due to the flexibility afforded by Rydberg physics.

We first analyze a frozen Rydberg aggregate, in which the positions $R_1,\dots R_N\equiv
\boldsymbol{R}$ of the atoms are assumed fixed and treat the electronic excited
states of the system, where we use a simple model of $N$ two-level Rydberg atoms. We define atomic states $\ket{nl}$, with principal quantum number $n$ and angular momentum $l$, and concentrate on a lower Rydberg state $|s_i\rangle=\ket{ns}$ of energy $E_s$, and a higher Rydberg state
$|p_i\rangle=\ket{np}$ of energy $E_p$. Here the index $i=1,\cdots,N$ labels the atoms. We consider the dynamics in the subspace spanned by the states $|\pi_i\rangle=|s_1,\cdots,s_{i-1},p_i,s_{i+1}\cdots,s_N\rangle$, where only the $i$-th atom is
in the $(np)$-state and all others are in the $(ns)$-state. In this subspace we
write the relevant Hamiltonian as
\begin{equation}\label{HGeneral}
H_{ex}=\sum_i E_i(\boldsymbol{R})\ket{\pi_i}\bra{\pi_i}+\sum_{ij}J_{ij}(\boldsymbol{R}) \ket{\pi_i}\bra{\pi_j},
\end{equation}
where $J_{ij}(R_i,R_j)\propto 1/|R_i-R_j|^3$ is the resonant dipole-dipole interaction. It is responsible for transfer of excitation between states $\ket{\pi_i}$ and $\ket{\pi_j}$.
The diagonal energies $E_i$ contain the off-resonant vdW interactions.
They can approximately written as
$E_i=E_0 + E_i^{(vdW)}({\mathbf{R}}) $, with $E_0= E_p+ (N-1) E_s $, and
$E_i^{(vdW)}({\mathbf{R}})\approx h \sum_{\ell \ne i } C^{sp}_6 /R_{i\ell}^6+
(h/2)\sum_{j\ne i}\sum_{\ell \ne i,j } C^{ss}_6 /R_{j\ell}^6$, where
$R_{i\ell}=|R_i-R_{\ell}|$, and $h$ is the Planck constant. Here $C^{ss}_6$
and $C^{sp}_6$ denote the $C_6$ coefficients for the vdW interaction between
two atoms in the $s$-state, and one atom in $s$ and the other in $p$,
respectively. With \bref{HGeneral} we model only electronic quantum states and
couplings that are relevant for the results presented here. They emerge in an
essential state picture from a more complete diagonalisation of the atomic
interaction Hamiltonian. For a discussion of three-body effects see \cite{Pohl2009}.

Note that  the magnitude and the sign of the $C_6$ coefficients depends on the chosen states (and atomic species).
The resonant interaction $J_{ij}$ depends in addition on the magnetic quantum number \cite{RoHeTo04_042703}  which leads to an anisotropic spatial interaction.
For the following considerations (where we restrict ourselves to one dimensional geometries) we ignore this anisotropy and write   $J_{ij}(R_i,R_j)= hC_3/|R_{ij}|^3$, where $C_3$ can be both positive or negative.

\begin{figure}[t]
\centerline{\epsfxsize=\columnwidth \epsfbox{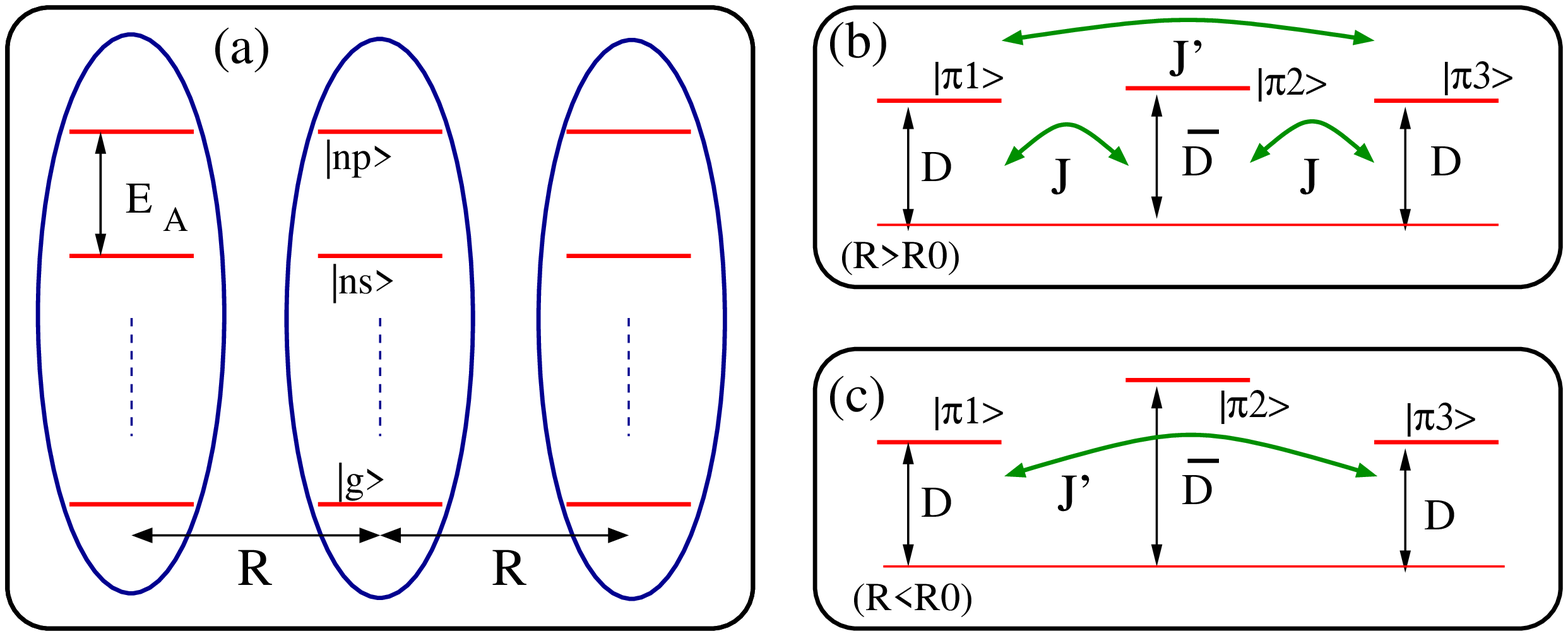}}
\caption{(Color online) (a) Single body states of three Rydberg atoms are illustrated schematically. We show the $|ns\rangle$ and $|np\rangle$ states with
  the transition energy $E_A=E_p-E_s$, and the absolute
  ground state $|g\rangle$. (b) The states $|\pi_i\rangle$, with $(i=1,2,3)$, are presented. For large interatomic distance relative to the crossover distance, $R>R_0$, the
  three $|\pi_i\rangle$ states are close to resonance, $D\approx \bar{D}$, and energy transfer is
  permitted among them with transfer parameters $J$ and $J'$. (c) For small interatomic distance, $R_0>R$, the states $|\pi_1\rangle$ and $|\pi_3\rangle$ have the same vdW shift of $D$, but $|\pi_2\rangle$ experiences a different shift $\bar{D}$ with $\bar{D}\gg D$. Hence,
  the $|\pi_2\rangle$ state is no longer resonant with $\ket{\pi_{1,3}}$. Now energy
  transfer is allowed only among the $|\pi_1\rangle$ and $|\pi_3\rangle$
  states with the transfer parameter $J'$.
\label{skematic}}
\end{figure}

 Diagonalization of the Hamiltonian (\ref{HGeneral}) for fixed positions
 $\boldsymbol{R}$ leads to (adiabatic) eigenstates
 $\ket{\psi_k(\boldsymbol{R})}=\sum_{i=1}^N c_{k}^i(\boldsymbol{R})
 \ket{\pi_i} $ and eigenenergies $U_k(\boldsymbol{R})$. To illustrate the basic electronic structure,
we discuss the case of a finite linear chain
of three Rydberg
atoms with equal spacing $R$ between nearest neighbors. The above excitonic Hamiltonian (\ref{HGeneral}) can then be written in the basis $\{\ket{\pi_1}$,$\ket{\pi_2}$,$\ket{\pi_3}\}$, as
\begin{eqnarray}\label{HamThree}
H_{ex}=E_0+\left(\begin{tabular}{ccc}
$D$ & $J$   & $J'$ \\
$J$   & $\bar{D}$ & $J$ \\
$J'$   & $J$  & $D$
\end{tabular}
\right),
\end{eqnarray}
where $J=\frac{hC_3}{R^3}$, and $J^{\prime}=J/8$, with
\begin{equation}\label{ParmThree}
D=\frac{h(C_6^{sp}+C_6^{ss})}{R^6}+D',\ \bar{D}=\frac{2hC_6^{sp}}{R^6}+\bar{D}',
\end{equation}
where $D'=hC_6^{sp}/(64R^6)$ and $\bar{D}'=hC_6^{ss}/(64R^6)$. Note that
$D'$, $\bar{D}'$ and $J'$ are small and can often be neglected to a good approximation.

From Hamiltonian (\ref{HamThree}) it is apparent that the relevance of the
relative shift $(\bar{D}-D)$ in the site energy of atom 2 is determined by the
ratio of $J\sim R^{-3}$ and $(\bar{D}-D)\sim R^{-6}$, which depend differently on the distance between
the atoms. In particular we consider two cases: large and small interatomic distances relative to the
`crossover distance', $R_0\sim |(C_6^{sp}-C_6^{ss})/C_3|^{1/3}$, where the
magnitude of $J$ becomes of the order of $(\bar{D}-D)$. For $R>R_0$ the three states $|\pi_i\rangle$ with
$(i=1,2,3)$  are close to resonance and energy transfer is
possible among them, as presented in \frefp{skematic}{b}. Hence the three states can
be coherently mixed to form excitonic states. In the case of $R<R_0$ the two states $|\pi_1\rangle$
and $|\pi_3\rangle$ are in resonance but the state $|\pi_2\rangle$ is shifted
off resonance due to vdW interactions, as sketched in \frefp{skematic}{c}. Now energy
transfer is possible mainly among the states $|\pi_1\rangle$
and $|\pi_3\rangle$, which combine into excitonic states, while the state
$|\pi_2\rangle$ remains a localized state. Thus, in this case the small $J'$
cannot be neglected.

This can also clearly be seen in our numerical calculations. For these we choose parameters $C_3=-1.16\ \mbox{GHz}(\mu
m)^3$, $C_6^{ss} = 47\ \mbox{MHz}(\mu m)^6$, appropriate for a principal quantum number $n=30$ \cite{Singer2005}, and $C_6^{sp}\approx +282\ \mbox{MHz}(\mu m)^6$. The latter might require
external modification of interaction strengths with the use of F{\"o}rster
resonances \cite{Altiere2011,nipper:foerster}.

\begin{figure}
\centerline{\epsfxsize=3cm \epsfbox{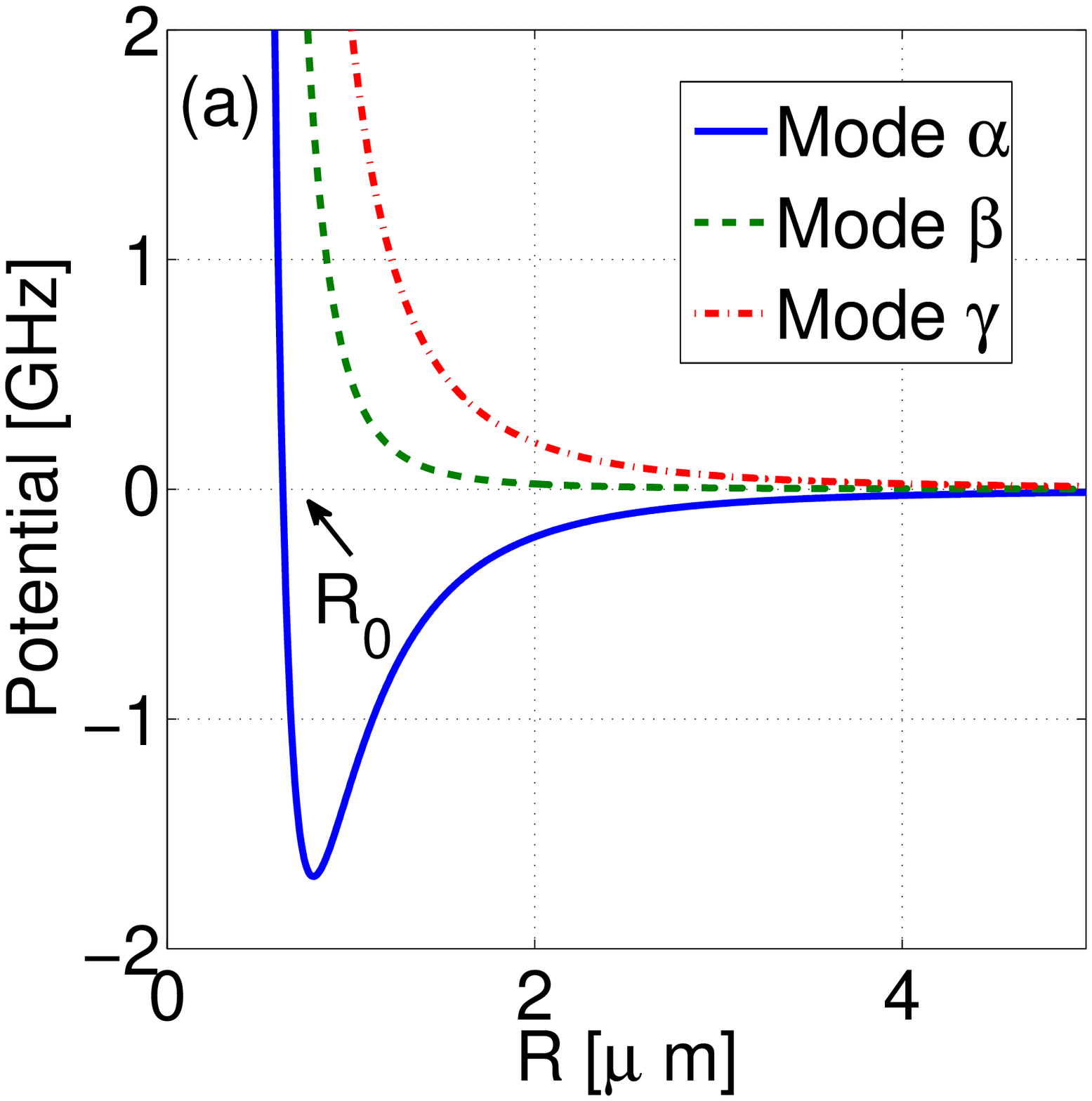}\epsfxsize=3cm \epsfbox{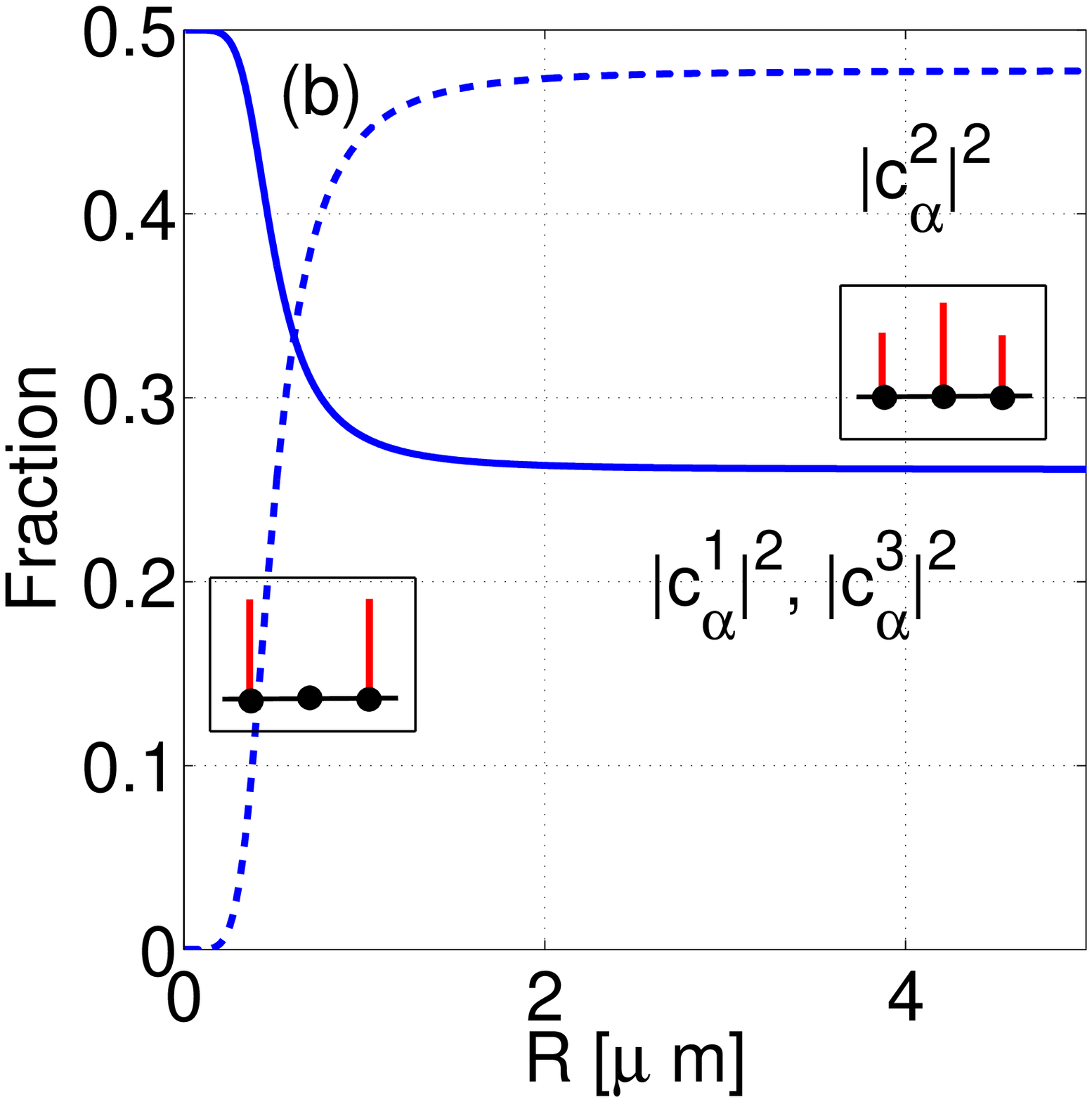}\epsfxsize=3cm \epsfbox{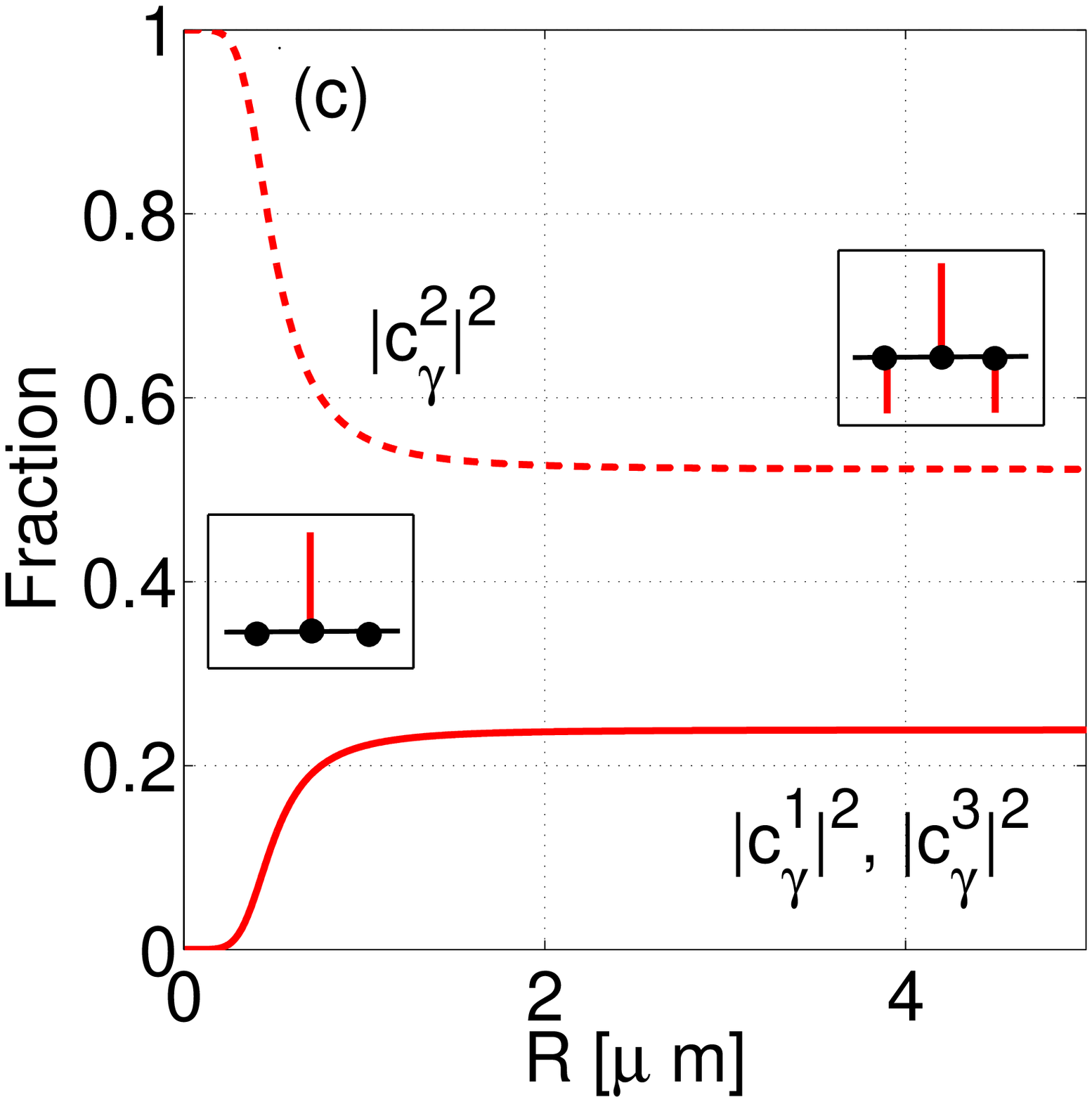}}
\caption{(Color online) (a) The three collective potentials, $U_k$, as a function
  of the interatomic distance $R$. (b-c) The fractions $|c^i_k|^2$ as a function of the interatomic distance $R$. (b) In
  the first $(k=\alpha)$, and (c) the third $(k=\gamma)$ collective states. The insets
  include schematically the mixing amplitudes at the three $|\pi_i\rangle$ states for the first
  and the third collective modes at large distances, $R>R_0$, and small distances, $R<R_0$.
  \label{potentials}
  }
\end{figure}

In \frefp{potentials}{a} we plot $U_k(R)$, for the three collective potentials (obtained by numerical diagonalisation of the Hamiltonian (\ref{HamThree})) relative
to $E_0/h$, as a function of the
interatomic distance $R$. The lowest potential has a minimum around $R\approx
0.8\ \mu m$, which leads to a bound state of the atoms. In
\frefp{potentials}{b-c} we plot the fraction of
each $|\pi_i\rangle$ state in the three collective states,
with $(k=\alpha,\beta,\gamma)$, that is $|c_k^i|^2$, as a function of $R$. In
mode $\alpha$, for small distances the
excitation is concentrated in the $|\pi_1\rangle$ and $|\pi_3\rangle$ states
and the state $|\pi_2\rangle$ is almost not excited. For large distances, a collective state is obtained with half
the excitation fraction on the $|\pi_2\rangle$ state and a quarter on each of
the $|\pi_1\rangle$ and $|\pi_3\rangle$ states. In mode $\beta$ (not shown) the $|\pi_2\rangle$ state is never involved in the formation of
the collective state. In mode $\gamma$ for small $R$ the excitation is almost entirely localized on the
$|\pi_2\rangle$ state, the $|\pi_1\rangle$ and $|\pi_3\rangle$ states are not excited. For
large $R$ the $\gamma$ state has the same population distribution as the $\alpha$ state. The amplitudes of the three $|\pi_i\rangle$ states
are plotted schematically in the insets of \frefp{potentials}{b-c} for large and small $R$.

The situation is different when excitons are formed via dipole-dipole interactions
involving ultracold ground- and low excited states in an optical lattice \cite{Zoubi2007,Zoubi2010,Zoubi2013},
or Rydberg states at larger separations than considered here
\cite{wuester:cradle,moebius:cradle}. Then the vdW shifts discussed above are
usually negligible.

After having established the effect of vdW interactions on the static
properties of the exciton states in a chain of Rydberg atoms at fixed
positions, we now consider dynamic properties of such a chain: a flexible
Rydberg aggregate \cite{leonhardt:switching}. This is done by augmenting the
total Hamiltonian of the problem with its kinetic energy part
$\sub{\hat{H}}{kin}=\sum_i P_i^2/(2M)$, where $P_i$ is the momentum of the
$i$'th atom, and $M$ its mass (for the examples shown, we choose Li). The combined treatment of exciton dynamics and atomic motion is involved quantum mechanically, but can be treated very well using
 Tully's quantum-classical description \cite{tully:hopping,cenap:motion,wuester:cradle,moebius:cradle}. In Tully's method, the internal electronic degrees of freedom are treated
quantum mechanically, while the external position degrees of freedom of the atom are treated classically. The $i$'th atom experiences a force  $F_{k}^i=-\nabla_{R_i}
U_{k}(\bv{R})$ according to one specific Born-Oppenheimer (BO) surface
$U_{k}$. This surface corresponds to the $k$'th eigenvector of the electronic
Hamiltonian $\sub{H}{ex}$. Due to these forces we obtain
time-dependent atomic trajectories $R_i(t)$ from the equation of motion $M\frac{d^2R_i}{dt^2}=F_k^i$, and excitation amplitudes
$c^i_k(t)$ from $i\partial_t
\ket{\Psi_k(t)}=\sub{\hat{H}}{ex}(\bv{R}(t))\ket{\Psi_k(t)}$, where $\ket{\Psi_k(t)}=\sum_i c^i_k(t)\ket{\pi_i}$. Stochastic non-adiabatic switches of the BO surface $k$ are possible~\cite{tully:hopping},
but occur in a negligible fraction ($<1\%$) of trajectories for all cases shown below.

We consider one-dimensionally confined Rydberg aggregates.
First, we discuss a trimer aggregate (i.e.$~$three atoms), where we
focus on the potential surface which stems from the combination of a repulsive
vdW interaction and an attractive `pure dipole-dipole surface', as in the $k=\alpha$ state of \frefp{potentials}{a}, since this may allow  a stable configuration due to the potential minimum.

\begin{figure}
\centerline{\epsfxsize=0.8\columnwidth \epsfbox{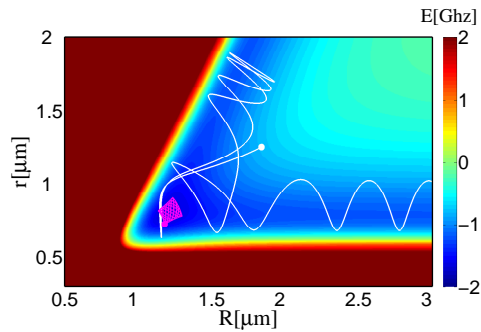}}
\caption{(Color online) Energy landscape of Rydberg trimer aggregate \cite{footnote:explaining:Jacobicoordinates}. Lines show a dissociating (white) and a stable (magenta) quantum classical trajectory. Big dots mark the starting position.
\label{energylandscape}}
\end{figure}

For a symmetric trimer the forces on the atoms can be solved analytically, in the general case the corresponding energy landscape has the shape shown in \fref{energylandscape} \cite{footnote:explaining:Jacobicoordinates}.
It allows stable bound trimer-aggregates in the global minimum around
$r\approx0.8\mu$m, $R\approx1.2\mu$m, as seen from the magenta trajectory.

\begin{figure}
\centerline{\epsfxsize=\columnwidth \epsfbox{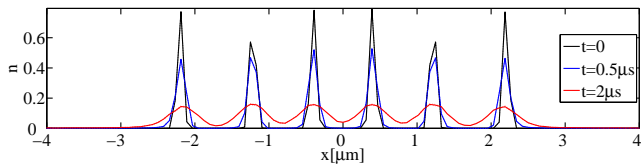}}
\caption{(Color online) Total atom density $n$ [arb.u.] in vdW stabilized Rydberg $6$-mer for three selected times.
\label{aggregate}}
\end{figure}

The binding on the lowest Born-Oppenheimer surface can be extended to larger aggregates, as shown in \fref{aggregate} for the case of six atoms. Atomic initial position and momentum are randomly distributed, according to the quantum ground state of an initially harmonically confined particle. We then bin atomic positions $R_i(t)$ to obtain a total atomic density $n(x,t)$. It shows partial dissociation, but a signature of the six-atom aggregate clearly remains visible. This stabilisation crucially requires the modification of exciton states by the vdW shifts discussed earlier. Without them, the exciton wave function has insufficient amplitude on the outermost atoms and attraction of these atoms becomes too weak. We choose a Rydberg state $n=30$ to maximize the product of self-trapping frequencies in the aggregate and system life-time ($3\mu$s for the case shown). This empirically favors smaller principal quantum numbers.

\begin{figure}
\centerline{\epsfxsize=0.8\columnwidth \epsfbox{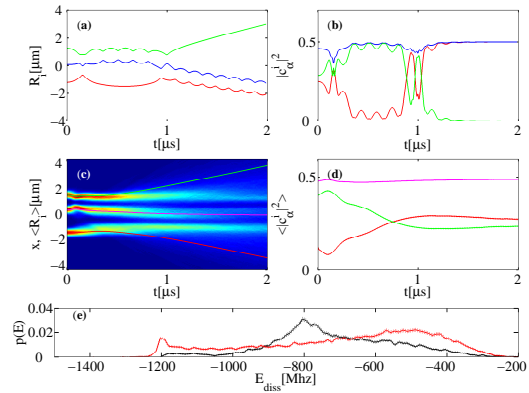}}
\caption{(Color online) Rydberg aggregate dissociation for (a,b) symmetric and (c,d) asymmetric configurations. (a) Atomic positions $R_i$ for single dissociating trajectory.
  (b) Excitation amplitudes $|c^i_{\alpha}(t)|^2$ for single dissociating trajectory. (c)
  Total atomic density $n$ from trajectory average, overlayed are the mean
  position of the three atoms. (d) Mean excitation amplitudes
  $\overline{|c^i_{\alpha}(t)|^2}$ on the three sites. (e) Energy distribution
  of the final-state dimer (black solid) and initial energy (red solid) for
  the same case as (c). The dotted lines indicate one standard error, and
  are nearly indistinguishable from the solid ones.
\label{dissociation}}
\end{figure}

Beside stable configuration one can prepare the system initially in
configurations that exhibit a more complicated dynamics, e.g.~the white
trajectory shown in \fref{energylandscape}. Details for this trajectory
are displayed in \frefp{dissociation}{a-b}. We can access regimes in which the trimer first undergoes breathing oscillations with excitation transfer between the sites, only to finally dissociate into a dimer and a free atom at a dissociation time $\sub{t}{diss}\approx1\mu$s, significantly faster than the excited system life-time of about $\sub{\tau}{life}=6{\mu}$s \cite{beterov:BBR}.

The dissociation can also be studied in a trajectory average, where we
obtain the pictures shown in \frefp{dissociation}{c-e}. Here, we consider a
chain which is slightly perturbed from a symmetric configuration. The position of each atom
is initially Gaussian distributed around $x=-1.5\mu$m,  $0.3\mu$m,  and
$1.5\mu$m. In the dynamics we can resolve some coherent motion of the chain in the total atomic density
(\frefp{dissociation}{c}), accompanied by excitation redistribution (\frefp{dissociation}{d}). Later
dissociation events that are qualitatively as in \frefp{dissociation}{a,b} smear out
the picture. Details of dissociation events depend strongly on the precise classical initial state, so that the
multi-trajectory simulation shows a quite broad distribution of the final
dimer energy \cite{footnote:dimer_energy}. The smallest dimer potential energy
is $\sub{U}{min}\approx-1.2$ GHz and the largest initial total energy about
$\sub{U}{0,tot}\approx-200$ MHz, explaining the bounds of the energy spectrum
in \frefp{dissociation}{e}.

One dimensional confinement of Rydberg atoms may soon be possible optically~\cite{li:lightatomentangle,rick:Rydberglattice}, hence our system provides a platform for the direct visualization of wave-packet dynamics in an analog of molecular-dissociation processes, with the use of state-selective and high-resolution Rydberg atom monitoring schemes~\cite{schwarzkopf:correlations,olmos:amplification,guenter:EIT,guenter:EITdetscience}.

In conclusion, we addressed the formation of excitons in chains of Rydberg
atoms with short separations, where vdW effect are shown to modify the
resonant dipole-dipole picture. The level of excitation localization depends
on the separation of atoms in the chain. We show that the combined action of
vdW and resonant dipole-dipole forces can lead to interesting effects in one-dimensional chains, such as the stabilization of larger Rydberg aggregates in an attractive exciton.
Even unstable parameter regimes show intriguing break-up dynamics, that can be
followed experimentally in quite some detail in the realm of ultra-cold
Rydberg physics. Moreover, using the HOMO and LUMO molecular configurations, similar exciton state modifications are expected for a cluster of organic molecules.

%\bibliography{VdWaggregates}

\begin{thebibliography}{37}
\expandafter\ifx\csname natexlab\endcsname\relax\def\natexlab#1{#1}\fi
\expandafter\ifx\csname bibnamefont\endcsname\relax
  \def\bibnamefont#1{#1}\fi
\expandafter\ifx\csname bibfnamefont\endcsname\relax
  \def\bibfnamefont#1{#1}\fi
\expandafter\ifx\csname citenamefont\endcsname\relax
  \def\citenamefont#1{#1}\fi
\expandafter\ifx\csname url\endcsname\relax
  \def\url#1{\texttt{#1}}\fi
\expandafter\ifx\csname urlprefix\endcsname\relax\def\urlprefix{URL }\fi
\providecommand{\bibinfo}[2]{#2}
\providecommand{\eprint}[2][]{\url{#2}}

\bibitem[{\citenamefont{Frenkel}(1930)}]{Fr30_198}
\bibinfo{author}{\bibfnamefont{J.}~\bibnamefont{Frenkel}},
  \bibinfo{journal}{Zeitschrift f{\"u}r Physik A}
  \textbf{\bibinfo{volume}{59}}, \bibinfo{pages}{198} (\bibinfo{year}{1930}).

\bibitem[{\citenamefont{Frenkel}(1931)}]{frenkel_exciton}
\bibinfo{author}{\bibfnamefont{J.}~\bibnamefont{Frenkel}},
  \bibinfo{journal}{Phys. Rev.} \textbf{\bibinfo{volume}{37}},
  \bibinfo{pages}{17} (\bibinfo{year}{1931}).

\bibitem[{\citenamefont{Davydov}(1971)}]{Davydov}
\bibinfo{author}{\bibfnamefont{S.}~\bibnamefont{Davydov}},
  \emph{\bibinfo{title}{Theory of Molecular Excitons}}
  (\bibinfo{publisher}{Plenum, NY}, \bibinfo{year}{1971}).

\bibitem[{\citenamefont{Agranovich}(2009)}]{Agranovich}
\bibinfo{author}{\bibfnamefont{V.~M.} \bibnamefont{Agranovich}},
  \emph{\bibinfo{title}{Excitations in Organic Solids}}
  (\bibinfo{publisher}{Oxford, UK}, \bibinfo{year}{2009}).

\bibitem[{\citenamefont{Zoubi and La Rocca}(2005)}]{Zoubi2005}
\bibinfo{author}{\bibfnamefont{H.}~\bibnamefont{Zoubi}} \bibnamefont{and}
  \bibinfo{author}{\bibfnamefont{G.~C.} \bibnamefont{La Rocca}},
  \bibinfo{journal}{Phys. Rev. B} \textbf{\bibinfo{volume}{71}},
  \bibinfo{pages}{235316} (\bibinfo{year}{2005}).

\bibitem[{\citenamefont{Roden et~al.}(2011)\citenamefont{Roden, Eisfeld,
  Dvo\v{r}\'ak, B\"unermann, and Stienkemeier}}]{RoEiDv11_054907}
\bibinfo{author}{\bibfnamefont{J.}~\bibnamefont{Roden}},
  \bibinfo{author}{\bibfnamefont{A.}~\bibnamefont{Eisfeld}},
  \bibinfo{author}{\bibfnamefont{M.}~\bibnamefont{Dvo\v{r}\'ak}},
  \bibinfo{author}{\bibfnamefont{O.}~\bibnamefont{B\"unermann}},
  \bibnamefont{and}
  \bibinfo{author}{\bibfnamefont{F.}~\bibnamefont{Stienkemeier}},
  \bibinfo{journal}{Journal of Chemical Physics}
  \textbf{\bibinfo{volume}{134}}, \bibinfo{pages}{054907}
  (\bibinfo{year}{2011}).

\bibitem[{\citenamefont{M\"{u}ller
  et~al.}(2013{\natexlab{a}})\citenamefont{M\"{u}ller, Paulheim, Marquardt, and
  Sokolowski}}]{MPaMa13_064703}
\bibinfo{author}{\bibfnamefont{M.}~\bibnamefont{M\"{u}ller}},
  \bibinfo{author}{\bibfnamefont{A.}~\bibnamefont{Paulheim}},
  \bibinfo{author}{\bibfnamefont{C.}~\bibnamefont{Marquardt}},
  \bibnamefont{and}
  \bibinfo{author}{\bibfnamefont{M.}~\bibnamefont{Sokolowski}},
  \bibinfo{journal}{The Journal of Chemical Physics}
  \textbf{\bibinfo{volume}{138}}, \bibinfo{pages}{064703}
  (\bibinfo{year}{2013}{\natexlab{a}}).

\bibitem[{\citenamefont{M\"{u}ller
  et~al.}(2013{\natexlab{b}})\citenamefont{M\"{u}ller, Paulheim, Eisfeld, and
  Sokolowski}}]{MPaMa13_044302}
\bibinfo{author}{\bibfnamefont{M.}~\bibnamefont{M\"{u}ller}},
  \bibinfo{author}{\bibfnamefont{A.}~\bibnamefont{Paulheim}},
  \bibinfo{author}{\bibfnamefont{A.}~\bibnamefont{Eisfeld}}, \bibnamefont{and}
  \bibinfo{author}{\bibfnamefont{M.}~\bibnamefont{Sokolowski}},
  \bibinfo{journal}{The Journal of Chemical Physics}
  \textbf{\bibinfo{volume}{139}}, \bibinfo{pages}{044302}
  (\bibinfo{year}{2013}{\natexlab{b}}).

\bibitem[{\citenamefont{Robicheaux et~al.}(2004)\citenamefont{Robicheaux,
  Hern\'{a}ndez, Top\c{c}u, and Noordam}}]{RoHeTo04_042703}
\bibinfo{author}{\bibfnamefont{F.}~\bibnamefont{Robicheaux}},
  \bibinfo{author}{\bibfnamefont{J.~V.} \bibnamefont{Hern\'{a}ndez}},
  \bibinfo{author}{\bibfnamefont{T.}~\bibnamefont{Top\c{c}u}},
  \bibnamefont{and} \bibinfo{author}{\bibfnamefont{L.~D.}
  \bibnamefont{Noordam}}, \bibinfo{journal}{Physical Review A}
  \textbf{\bibinfo{volume}{70}}, \bibinfo{pages}{042703}
  (\bibinfo{year}{2004}).

\bibitem[{\citenamefont{Ates et~al.}(2008)\citenamefont{Ates, Eisfeld, and
  Rost}}]{cenap:motion}
\bibinfo{author}{\bibfnamefont{C.}~\bibnamefont{Ates}},
  \bibinfo{author}{\bibfnamefont{A.}~\bibnamefont{Eisfeld}}, \bibnamefont{and}
  \bibinfo{author}{\bibfnamefont{J.~M.} \bibnamefont{Rost}},
  \bibinfo{journal}{New J. Phys.} \textbf{\bibinfo{volume}{10}},
  \bibinfo{pages}{045030} (\bibinfo{year}{2008}).

\bibitem[{\citenamefont{Saffman et~al.}(2010)\citenamefont{Saffman, Walker, and
  Molmer}}]{Saffman}
\bibinfo{author}{\bibfnamefont{M.}~\bibnamefont{Saffman}},
  \bibinfo{author}{\bibfnamefont{T.~G.} \bibnamefont{Walker}},
  \bibnamefont{and} \bibinfo{author}{\bibfnamefont{K.}~\bibnamefont{Molmer}},
  \bibinfo{journal}{Rev. Mod. Phys.} \textbf{\bibinfo{volume}{82}},
  \bibinfo{pages}{2313} (\bibinfo{year}{2010}).

\bibitem[{\citenamefont{Gallagher and Pillet}(2008)}]{GallagherA}
\bibinfo{author}{\bibfnamefont{T.~F.} \bibnamefont{Gallagher}}
  \bibnamefont{and} \bibinfo{author}{\bibfnamefont{P.}~\bibnamefont{Pillet}},
  \bibinfo{journal}{Ad. At. Mol. Opt. Phys.} \textbf{\bibinfo{volume}{56}},
  \bibinfo{pages}{161} (\bibinfo{year}{2008}).

\bibitem[{\citenamefont{Bettelli et~al.}(2013)\citenamefont{Bettelli, Maxwell,
  Fernholz, Adams, Lesanovsky, and Ates}}]{cenap:emergingaggregate}
\bibinfo{author}{\bibfnamefont{S.}~\bibnamefont{Bettelli}},
  \bibinfo{author}{\bibfnamefont{D.}~\bibnamefont{Maxwell}},
  \bibinfo{author}{\bibfnamefont{T.}~\bibnamefont{Fernholz}},
  \bibinfo{author}{\bibfnamefont{C.~S.} \bibnamefont{Adams}},
  \bibinfo{author}{\bibfnamefont{I.}~\bibnamefont{Lesanovsky}},
  \bibnamefont{and} \bibinfo{author}{\bibfnamefont{C.}~\bibnamefont{Ates}},
  \bibinfo{journal}{Phys. Rev. A} \textbf{\bibinfo{volume}{88}},
  \bibinfo{pages}{043436} (\bibinfo{year}{2013}).

\bibitem[{\citenamefont{Kiffner et~al.}(2012)\citenamefont{Kiffner, Park, Li,
  and Gallagher}}]{kiffner:dipdipdimer}
\bibinfo{author}{\bibfnamefont{M.}~\bibnamefont{Kiffner}},
  \bibinfo{author}{\bibfnamefont{H.}~\bibnamefont{Park}},
  \bibinfo{author}{\bibfnamefont{W.}~\bibnamefont{Li}}, \bibnamefont{and}
  \bibinfo{author}{\bibfnamefont{T.~F.} \bibnamefont{Gallagher}},
  \bibinfo{journal}{Phys. Rev. A} \textbf{\bibinfo{volume}{86}},
  \bibinfo{pages}{031401(R)} (\bibinfo{year}{2012}).

\bibitem[{\citenamefont{Kiffner
  et~al.}(2013{\natexlab{a}})\citenamefont{Kiffner, Li, and
  Jaksch}}]{kiffner:spinorbitcoupl}
\bibinfo{author}{\bibfnamefont{M.}~\bibnamefont{Kiffner}},
  \bibinfo{author}{\bibfnamefont{W.}~\bibnamefont{Li}}, \bibnamefont{and}
  \bibinfo{author}{\bibfnamefont{D.}~\bibnamefont{Jaksch}},
  \bibinfo{journal}{Phys. Rev. Lett.} \textbf{\bibinfo{volume}{110}},
  \bibinfo{pages}{170402} (\bibinfo{year}{2013}{\natexlab{a}}).

\bibitem[{\citenamefont{Kiffner
  et~al.}(2013{\natexlab{b}})\citenamefont{Kiffner, Li, and
  Gallagher}}]{kiffner:nonabelian}
\bibinfo{author}{\bibfnamefont{M.}~\bibnamefont{Kiffner}},
  \bibinfo{author}{\bibfnamefont{W.}~\bibnamefont{Li}}, \bibnamefont{and}
  \bibinfo{author}{\bibfnamefont{T.~F.} \bibnamefont{Gallagher}},
  \bibinfo{journal}{J. Phys. B: At. Mol. Opt. Phys.}
  \textbf{\bibinfo{volume}{46}}, \bibinfo{pages}{134008}
  (\bibinfo{year}{2013}{\natexlab{b}}).

\bibitem[{\citenamefont{Beterov et~al.}(2009)\citenamefont{Beterov, Ryabtsev,
  Tretyakov, and Entin}}]{beterov:BBR}
\bibinfo{author}{\bibfnamefont{I.~I.} \bibnamefont{Beterov}},
  \bibinfo{author}{\bibfnamefont{I.~I.} \bibnamefont{Ryabtsev}},
  \bibinfo{author}{\bibfnamefont{D.~B.} \bibnamefont{Tretyakov}},
  \bibnamefont{and} \bibinfo{author}{\bibfnamefont{V.~M.} \bibnamefont{Entin}},
  \bibinfo{journal}{Phys. Rev. A} \textbf{\bibinfo{volume}{79}},
  \bibinfo{pages}{052504} (\bibinfo{year}{2009}).

\bibitem[{\citenamefont{W{\"u}ster et~al.}(2010)\citenamefont{W{\"u}ster, Ates,
  Eisfeld, and Rost}}]{wuester:cradle}
\bibinfo{author}{\bibfnamefont{S.}~\bibnamefont{W{\"u}ster}},
  \bibinfo{author}{\bibfnamefont{C.}~\bibnamefont{Ates}},
  \bibinfo{author}{\bibfnamefont{A.}~\bibnamefont{Eisfeld}}, \bibnamefont{and}
  \bibinfo{author}{\bibfnamefont{J.~M.} \bibnamefont{Rost}},
  \bibinfo{journal}{Phys. Rev. Lett.} \textbf{\bibinfo{volume}{105}},
  \bibinfo{pages}{053004} (\bibinfo{year}{2010}).

\bibitem[{\citenamefont{M{\"o}bius et~al.}(2011)\citenamefont{M{\"o}bius,
  W{\"u}ster, Ates, Eisfeld, and Rost}}]{moebius:cradle}
\bibinfo{author}{\bibfnamefont{S.}~\bibnamefont{M{\"o}bius}},
  \bibinfo{author}{\bibfnamefont{S.}~\bibnamefont{W{\"u}ster}},
  \bibinfo{author}{\bibfnamefont{C.}~\bibnamefont{Ates}},
  \bibinfo{author}{\bibfnamefont{A.}~\bibnamefont{Eisfeld}}, \bibnamefont{and}
  \bibinfo{author}{\bibfnamefont{J.~M.} \bibnamefont{Rost}},
  \bibinfo{journal}{J. Phys. B: At. Mol. Opt. Phys.}
  \textbf{\bibinfo{volume}{44}}, \bibinfo{pages}{184011}
  (\bibinfo{year}{2011}).

\bibitem[{\citenamefont{Leonhardt et~al.}(2013)\citenamefont{Leonhardt,
  W{\"u}ster, and Rost}}]{leonhardt:switching}
\bibinfo{author}{\bibfnamefont{K.}~\bibnamefont{Leonhardt}},
  \bibinfo{author}{\bibfnamefont{S.}~\bibnamefont{W{\"u}ster}},
  \bibnamefont{and} \bibinfo{author}{\bibfnamefont{J.}~\bibnamefont{Rost}}
  (\bibinfo{year}{2013}), \eprint{(in preparation)}.

\bibitem[{\citenamefont{Pohl and Berman}(2009)}]{Pohl2009}
\bibinfo{author}{\bibfnamefont{T.}~\bibnamefont{Pohl}} \bibnamefont{and}
  \bibinfo{author}{\bibfnamefont{P.~R.} \bibnamefont{Berman}},
  \bibinfo{journal}{Phys. Rev. Lett.} \textbf{\bibinfo{volume}{102}},
  \bibinfo{pages}{013004} (\bibinfo{year}{2009}).

\bibitem[{\citenamefont{Singer et~al.}(2005)\citenamefont{Singer, Stanojevic,
  Weidem{\"u}ller, and {C\^ot\'e}}}]{Singer2005}
\bibinfo{author}{\bibfnamefont{K.}~\bibnamefont{Singer}},
  \bibinfo{author}{\bibfnamefont{J.}~\bibnamefont{Stanojevic}},
  \bibinfo{author}{\bibfnamefont{M.}~\bibnamefont{Weidem{\"u}ller}},
  \bibnamefont{and}
  \bibinfo{author}{\bibfnamefont{R.}~\bibnamefont{{C\^ot\'e}}},
  \bibinfo{journal}{J. Phys. B: At. Mol. Opt. Phys.}
  \textbf{\bibinfo{volume}{38}}, \bibinfo{pages}{{S295}}
  (\bibinfo{year}{2005}).

\bibitem[{\citenamefont{Altiere et~al.}(2011)\citenamefont{Altiere, Fahey,
  Noel, Smith, and Carroll}}]{Altiere2011}
\bibinfo{author}{\bibfnamefont{E.}~\bibnamefont{Altiere}},
  \bibinfo{author}{\bibfnamefont{D.~P.} \bibnamefont{Fahey}},
  \bibinfo{author}{\bibfnamefont{M.~W.} \bibnamefont{Noel}},
  \bibinfo{author}{\bibfnamefont{R.~J.} \bibnamefont{Smith}}, \bibnamefont{and}
  \bibinfo{author}{\bibfnamefont{T.~J.} \bibnamefont{Carroll}},
  \bibinfo{journal}{Phys. Rev. A} \textbf{\bibinfo{volume}{84}},
  \bibinfo{pages}{053431} (\bibinfo{year}{2011}).

\bibitem[{\citenamefont{Nipper et~al.}(2012)\citenamefont{Nipper, Balewski,
  Krupp, Butscher, L{\"o}w, and Pfau}}]{nipper:foerster}
\bibinfo{author}{\bibfnamefont{J.}~\bibnamefont{Nipper}},
  \bibinfo{author}{\bibfnamefont{J.~B.} \bibnamefont{Balewski}},
  \bibinfo{author}{\bibfnamefont{A.~T.} \bibnamefont{Krupp}},
  \bibinfo{author}{\bibfnamefont{B.}~\bibnamefont{Butscher}},
  \bibinfo{author}{\bibfnamefont{R.}~\bibnamefont{L{\"o}w}}, \bibnamefont{and}
  \bibinfo{author}{\bibfnamefont{T.}~\bibnamefont{Pfau}},
  \bibinfo{journal}{Phys. Rev. Lett.} \textbf{\bibinfo{volume}{108}},
  \bibinfo{pages}{113001} (\bibinfo{year}{2012}).

\bibitem[{\citenamefont{Zoubi and Ritsch}(2007)}]{Zoubi2007}
\bibinfo{author}{\bibfnamefont{H.}~\bibnamefont{Zoubi}} \bibnamefont{and}
  \bibinfo{author}{\bibfnamefont{H.}~\bibnamefont{Ritsch}},
  \bibinfo{journal}{Phys. Rev. A} \textbf{\bibinfo{volume}{76}},
  \bibinfo{pages}{013817} (\bibinfo{year}{2007}).

\bibitem[{\citenamefont{Zoubi and Ritsch}(2010)}]{Zoubi2010}
\bibinfo{author}{\bibfnamefont{H.}~\bibnamefont{Zoubi}} \bibnamefont{and}
  \bibinfo{author}{\bibfnamefont{H.}~\bibnamefont{Ritsch}},
  \bibinfo{journal}{Europhys. Lett.} \textbf{\bibinfo{volume}{90}},
  \bibinfo{pages}{23001} (\bibinfo{year}{2010}).

\bibitem[{\citenamefont{Zoubi and Ritsch}(2013)}]{Zoubi2013}
\bibinfo{author}{\bibfnamefont{H.}~\bibnamefont{Zoubi}} \bibnamefont{and}
  \bibinfo{author}{\bibfnamefont{H.}~\bibnamefont{Ritsch}},
  \bibinfo{journal}{Ad. At. Mol. Opt. Phys.} \textbf{\bibinfo{volume}{62}},
  \bibinfo{pages}{171} (\bibinfo{year}{2013}).

\bibitem[{\citenamefont{Tully}(1990)}]{tully:hopping}
\bibinfo{author}{\bibfnamefont{J.~C.} \bibnamefont{Tully}},
  \bibinfo{journal}{J. Chem. Phys.} \textbf{\bibinfo{volume}{93}},
  \bibinfo{pages}{1061} (\bibinfo{year}{1990}).

\bibitem[{foo({\natexlab{a}})}]{footnote:explaining:Jacobicoordinates}
\bibinfo{note}{For ease of interpretation we use Jacobi-coordinates
  $r=R_2-R_1$, $R=R_3-R_2 -r/2$.}

\bibitem[{foo({\natexlab{b}})}]{footnote:dimer_energy}
\bibinfo{note}{We wait for the classical trajectory to have clearly
  dissociated, and then determine the kinetic plus potential energy of the
  dimer part of the system only.}

\bibitem[{\citenamefont{Li et~al.}(2013)\citenamefont{Li, Dudin, and
  Kuzmich}}]{li:lightatomentangle}
\bibinfo{author}{\bibfnamefont{L.}~\bibnamefont{Li}},
  \bibinfo{author}{\bibfnamefont{Y.~O.} \bibnamefont{Dudin}}, \bibnamefont{and}
  \bibinfo{author}{\bibfnamefont{A.}~\bibnamefont{Kuzmich}},
  \bibinfo{journal}{Nature} \textbf{\bibinfo{volume}{498}},
  \bibinfo{pages}{466} (\bibinfo{year}{2013}).

\bibitem[{\citenamefont{Mukherjee et~al.}(2011)\citenamefont{Mukherjee, Millen,
  Nath, Jones, and Pohl}}]{rick:Rydberglattice}
\bibinfo{author}{\bibfnamefont{R.}~\bibnamefont{Mukherjee}},
  \bibinfo{author}{\bibfnamefont{J.}~\bibnamefont{Millen}},
  \bibinfo{author}{\bibfnamefont{R.}~\bibnamefont{Nath}},
  \bibinfo{author}{\bibfnamefont{M.~P.~A.} \bibnamefont{Jones}},
  \bibnamefont{and} \bibinfo{author}{\bibfnamefont{T.}~\bibnamefont{Pohl}},
  \bibinfo{journal}{J. Phys. B: At. Mol. Opt. Phys.}
  \textbf{\bibinfo{volume}{33}}, \bibinfo{pages}{184010}
  (\bibinfo{year}{2011}).

\bibitem[{\citenamefont{Schwarzkopf et~al.}(2011)\citenamefont{Schwarzkopf,
  Sapiro, and Raithel}}]{schwarzkopf:correlations}
\bibinfo{author}{\bibfnamefont{A.}~\bibnamefont{Schwarzkopf}},
  \bibinfo{author}{\bibfnamefont{R.~E.} \bibnamefont{Sapiro}},
  \bibnamefont{and} \bibinfo{author}{\bibfnamefont{G.}~\bibnamefont{Raithel}},
  \bibinfo{journal}{Phys. Rev. Lett.} \textbf{\bibinfo{volume}{107}},
  \bibinfo{pages}{103001} (\bibinfo{year}{2011}).

\bibitem[{\citenamefont{Olmos et~al.}(2011)\citenamefont{Olmos, Li,
  Hofferberth, and Lesanovsky}}]{olmos:amplification}
\bibinfo{author}{\bibfnamefont{B.}~\bibnamefont{Olmos}},
  \bibinfo{author}{\bibfnamefont{W.}~\bibnamefont{Li}},
  \bibinfo{author}{\bibfnamefont{S.}~\bibnamefont{Hofferberth}},
  \bibnamefont{and}
  \bibinfo{author}{\bibfnamefont{I.}~\bibnamefont{Lesanovsky}},
  \bibinfo{journal}{Phys. Rev. A} \textbf{\bibinfo{volume}{84}},
  \bibinfo{pages}{041607(R)} (\bibinfo{year}{2011}).

\bibitem[{\citenamefont{G{\"u}nter et~al.}(2012)\citenamefont{G{\"u}nter,
  Robert-de-Saint-Vincent, Schempp, Hofmann, Whitlock, and
  Weidem{\"u}ller}}]{guenter:EIT}
\bibinfo{author}{\bibfnamefont{G.}~\bibnamefont{G{\"u}nter}},
  \bibinfo{author}{\bibfnamefont{M.} \bibnamefont{Robert-de-Saint-Vincent}},
  \bibinfo{author}{\bibfnamefont{H.}~\bibnamefont{Schempp}},
  \bibinfo{author}{\bibfnamefont{C.~S.} \bibnamefont{Hofmann}},
  \bibinfo{author}{\bibfnamefont{S.}~\bibnamefont{Whitlock}}, \bibnamefont{and}
  \bibinfo{author}{\bibfnamefont{M.}~\bibnamefont{Weidem{\"u}ller}},
  \bibinfo{journal}{Phys. Rev. Lett.} \textbf{\bibinfo{volume}{108}},
  \bibinfo{pages}{013002} (\bibinfo{year}{2012}).

\bibitem[{\citenamefont{G{\"u}nter et~al.}(2013)\citenamefont{G{\"u}nter,
  Schempp, de~Saint-Vincent, Gavryusev, Helmrich, Hofmann, Whitlock, and
  Weidem{\"u}ller}}]{guenter:EITdetscience}
\bibinfo{author}{\bibfnamefont{G.}~\bibnamefont{G{\"u}nter}},
  \bibinfo{author}{\bibfnamefont{H.}~\bibnamefont{Schempp}},
  \bibinfo{author}{\bibfnamefont{M.~R.} \bibnamefont{de~Saint-Vincent}},
  \bibinfo{author}{\bibfnamefont{V.}~\bibnamefont{Gavryusev}},
  \bibinfo{author}{\bibfnamefont{S.}~\bibnamefont{Helmrich}},
  \bibinfo{author}{\bibfnamefont{C.~S.} \bibnamefont{Hofmann}},
  \bibinfo{author}{\bibfnamefont{S.}~\bibnamefont{Whitlock}}, \bibnamefont{and}
  \bibinfo{author}{\bibfnamefont{M.}~\bibnamefont{Weidem{\"u}ller}},
  \bibinfo{journal}{Science} \textbf{\bibinfo{volume}{342}},
  \bibinfo{pages}{954} (\bibinfo{year}{2013}).

\end{thebibliography}

\end{document}